\documentclass{emulateapj}

\newcommand{\sigs}{$\sigma_{*}$}
\newcommand{\Vs}{$V_{*}$}
\newcommand{\sige}{$\sigma_{e}$}
\newcommand{\re}{$r_{e}$}
\newcommand{\Mb}{$M_{bul}$}
\newcommand{\Mbserel}{$M_{\bullet} - \sigma_{e}$ relation}
\newcommand{\Mbssrel}{$M_{\bullet} - \sigma_{*}$ relation}
\newcommand{\Mbh}{$M_{\bullet}$}

\newcommand{\asec}{$^{\prime \prime}$ }

\newcommand{\kms}{km~s$^{-1}$ }

\newcommand{\fwoiii}{$\rm FWHM_{[OIII]}$ }

\newcommand{\magx}{$\phantom{.}.\!\!\!^{\rm m}$}

\newcommand{\Cat}{Ca~II triplet}

\newcommand{\cencol}[1]{\multicolumn{1}{c}{#1}}
\newcommand\eol{\\}
\newcommand\extline{ \eol}

\begin{document}

\title{The Relationship Between Black Hole Mass and Velocity Dispersion 
in Seyfert 1 Galaxies}

\author{Charles H. Nelson\altaffilmark{1}}

\affil{Physics and Astronomy Department, Drake University, 2507
University Ave., \\ Des Moines, IA 50311, charles.nelson@drake.edu}


\author{Richard F. Green\altaffilmark{1,2}} 
\affil{National Optical Astronomy 
Observatory, \\ P. O. Box 26732, Tucson, AZ 85726}


\author{Gary Bower\altaffilmark{1}}
\affil{Space Telescope Science Institute,
3700 San Martin Drive, Baltimore, MD 21218}

\author{Karl Gebhardt}
\affil{Department of Astronomy, University of Texas at Austin, Austin, TX 78712}


\author{Donna Weistrop}
\affil{Physics Dept., University of Nevada, Las Vegas, Box 4002, 
4505 Maryland Pkwy., Las Vegas, NV 89154}

\altaffiltext{1}{Visiting Astronomer, Kitt Peak National Observatory,
National Optical Astronomy Observatory, which is operated by the
Association of Universities for Research in Astronomy, Inc. (AURA)
under cooperative agreement with the National Science Foundation.}

\altaffiltext{2}{Kitt Peak National Observatory and National Optical
Astronomy Observatory are operated by the Association of Universities
for Research in Astronomy, Inc. (AURA) under cooperative agreement
with the National Science Foundation.}

\begin{abstract}

Black hole masses in active galactic nuclei (AGN) are difficult to
measure using conventional dynamical methods, but can be determined
using the technique of reverberation mapping. However, it is important
to verify that the results of these different methods are equivalent.
This can be done indirectly, using scaling relations between the black
hole and the host galaxy spheroid.  For this purpose, we have obtained
new measurements of the bulge stellar velocity dispersion, \sigs, in
Seyfert 1 galaxies.  These are used in conjunction with the \Mbh\ --
\sigs \ relation to validate nuclear black hole masses, \Mbh, in
active galaxies determined through reverberation mapping.  We find
that Seyfert galaxies follow the same \Mbh\ -- \sigs \ relation as
non-active galaxies, indicating that reverberation mapping
measurements of \Mbh\ are consistent with those obtained using other
methods.  We also reconsider the relationship between bulge absolute
magnitude, \Mb, and black hole mass.  We find that Seyfert galaxies
are offset from non-active galaxies, but that the deviation can be
entirely understood as a difference in bulge luminosity, not black
hole mass; Seyfert hosts are brighter than normal galaxies for a given
value of their velocity dispersion, perhaps as a result of younger
stellar populations.

\end{abstract}

\keywords{galaxies: nuclei --- galaxies: Seyfert --- quasars:general ---
galaxies: kinematics and dynamics}

\section{Introduction}
\label{sec:intro}


An observational problem in research on active galactic nuclei (AGN)
is determining the masses of the central black holes.  Application of
stellar and gas dynamical techniques, commonly used for normal
galaxies \citep[e.g.][]{bower98,gebhardt03}, is difficult or
impossible in type 1 AGN due to the bright nucleus.  A potential
solution is the technique known as reverberation mapping
\citep{blandford82,np97}, which has been used to estimate black hole
masses for about 30 Seyfert galaxies and quasars
\citep{wpm99,kaspi2000}.  Reverberation mapping allows the radius of
the broad line region, $R_{BLR}$, to be determined by attributing the
time lag between variations in the continuum and line emission to
light travel time effects.  Assuming that the velocity width, $W$, of
the permitted emission lines is due to Keplerian motion in the black
hole potential, the mass can estimated by $M_{\bullet}\sim W^2
R_{BLR}/G$, where $G$ is the gravitational constant.  Thus, this
method may provide an opportunity to study the role of the nuclear
gravitational potential in large numbers of AGN.  But it has not yet
been demonstrated that reverberation mapping yields \Mbh \ values that
are consistent with those obtained from analysis of the stellar and
gas dynamics. Given the difficulty of a direct approach, indirect
methods must be used to compare these different techniques.  Our
purpose in this study is to compare reverberation mapping and stellar
dynamical measures of black hole mass using scaling relations between
the black hole and the host galaxy bulge.

The correlation between black hole mass and bulge absolute magnitude,
\Mb, \citep{Kor93,korrich}, can be used to compare results of these
different approaches.  However, comparisons of the \Mbh--\Mb\ relation
in active and normal galaxies have produced conflicting results.  Two
early studies \citep{ho99,wandel99} reported that the reverberation
mapping masses for AGN were smaller than the stellar dynamical masses
in normal galaxies of the same bulge luminosity.  This called into
question the validity of the reverberation mapping masses, suggesting
a systematic trend to underestimate \Mbh\ by a factor of $\sim 5$
\citep{ho99}.  Alternatively, \citet{wandel99} suggested that the
masses were accurate and that there were real differences in the black
hole masses of active and non-active galaxies of equal bulge
mass. More recently, \citet{wandel02}, revising his early assessment,
and \citet{mcluredunlop02}, using bulge magnitudes determined in most
of their sample from Hubble Space Telescope (HST) images and assuming
disk-like kinematics for the BLR, find that active galaxies follow the
same \Mbh\ -- \Mb \ relation as normal galaxies.

It is difficult to make a fair assessment of these differing results
since some of these studies have used bulge magnitudes derived from
the relation between bulge-to-disk ratio as a function of
morphological type formulated by \citet{simdevauc}, while others have
applied bulge-disk decomposition techniques to ground-based or HST
images.  Furthermore, bulges of Seyfert galaxies tend to be small ---
their typical effective radii in all but the nearest objects are a few
arcseconds or less.  This, combined with the effects of the bright nuclear
source in type 1 objects, which are of course the candidates for
reverberation mapping, makes ground based estimates of \Mb \
quite tricky.

The relationship between black hole mass and bulge stellar velocity
dispersion, \sigs, \citep{fer2000,geb2000a}, provides an alternative
and independent way to compare stellar dynamical and reverberation
mapping results.  In this relation, which is stronger
than the one using \Mb, \sigs\ is measured in apertures extending to
distances of a kiloparsec or so from the nucleus \citep[use apertures
extending to the bulge effective radius, \re, and denote their
measurement \sige]{geb2000a}. The goal is to standardize the region
over which the dispersion is determined and more importantly to
minimize the influence of the nuclear black hole on the stellar
kinematics. 

The small number of published \sigs\ measurements for type 1 AGN, has
until recently prevented a statistically significant
investigation. \citet{geb2000b} found no deviation for 7 Seyfert 1s
from the \Mbserel. The same conclusion was reached by \citet{nel2000},
using \fwoiii\ in place of \sigs, on the assumption that the forbidden
line kinematics in Seyfert galaxies are predominantly the result of
virial motion in the bulge gravitational potential
\citep[e.g.][]{nw96}.  Also, \citet{fer2001} obtained \sigs\
measurements for 6 Seyfert 1 galaxies and find these values to be
consistent with expectations from the \Mbssrel.

In this paper we present new measurements of \sigs\ in 14 Seyfert
1 galaxies to check the \Mbh \ values obtained from reverberation
mapping for consistency with the \Mbserel \ established by stellar dynamical
techniques.  Observations of the \Cat \ lines and velocity
dispersion measurements using the Fourier Correlation Quotient (FCQ)
are presented in section \ref{sec:obs}.  Bulge magnitude estimates for
11 of these obtained from Hubble Space Telescope archival data are
presented in section \ref{sec:bulmag}.  Notes on individual sample
members are presented in section \ref{sec:galnotes}. In section
\ref{sec:mbhsig} we discuss the results in the context of the 
\Mbssrel \ and in section \ref{sec:mbhmb} we revisit the
\Mbh\ -- \Mb\ relation to try to determine the origin of the discrepancies
noted above. 
Our results are summarized in section \ref{sec:summary}.

\begin{table}

\caption{\label{tab:s1sigma}}
\begin{centering}

\begin{tabular}{lcr @{$\pm$}lr @{$\pm$}ll}

\multicolumn{7}{c}{Velocity Dispersions in Seyfert 1 Galaxies}
\smallskip \\
\hline 
\vspace*{-5pt}
\extline
\cencol{Name} & Aperture  & $\sigma_*$ & $\epsilon$ & $cz$ & $\epsilon$ & 
\cencol{$\log M_{\bullet}^\ast$}\\
& ($\prime \prime$) & \multicolumn{2}{c}{(km/s)} & \multicolumn{2}{c}{(km/s)} & \cencol{($M_{\odot}$)} 
\smallskip \eol
\hline \hline
\vspace*{-5pt}
\extline
  NGC 3227 & 4.7 & ~~$ 136 $ & $4 $ & ~~~$  1187 $ & $  5  $ 
& ~~ $  7.56^{+0.14}_{-0.21}$ 
\vspace*{3pt} \\
  NGC 3516 & 4.3 & $ 181 $ & $5 $ & $  2657 $ & $  6  $ 
& ~~ $  7.23^{+0.08}_{-0.09}$ 
\vspace*{3pt}  \\
  NGC 4051 & 5.6 & $  89 $ & $3 $ & $   743 $ & $  6  $ 
& ~~ $ 6.15 ^{+0.32}_{-0.45}$ 
\vspace*{3pt}  \\
  NGC 4151 & 6.5 & $  97 $ & $3 $ & $  1020 $ & $  6  $ 
& ~~ $ 7.08 ^{+0.23}_{-0.38}$ 
\vspace*{3pt} \\
  NGC 4593 & 6.5 & $ 135 $ & $6 $ & $  2602 $ & $  7  $ 
& ~~ $  6.82^{+0.25}_{-0.67}$ 
\vspace*{3pt} \\
  NGC 5548 & 2.2 & $ 201 $ & $12 $ & $  5100 $ & $ 15  $ 
& ~~ $ 7.97 ^{+0.07}_{-0.07}$ 
\vspace*{3pt} \\
  NGC 7469 & 4.7 & $ 131 $ & $5 $ & $  4947 $ & $  7  $ 
& ~~ $ 6.88 ^{+0.30}_{-6.88}$ 
\vspace*{3pt} \\
  Mrk 79   & 4.1 & $ 130 $ & $12 $ & $  6540 $ & $  8  $ 
& ~~ $ 8.01 ^{+0.14}_{-0.35}$ 
\vspace*{3pt} \\
  Mrk 110  & 4.7 & $  91 $ & $25 $ & $ 10509 $ & $ 13  $ 
& ~~ $ 6.89 ^{+0.13}_{-0.21}$ 
\vspace*{3pt} \\
  Mrk 279  & 4.7 & $ 197 $ & $ 12 $ & $  9062 $ & $ 12  $ 
& ~~ $ 7.41 ^{+0.10}_{-0.10}$ 
\vspace*{3pt}  \\
  Mrk 335  & $-$ & \multicolumn{2}{c}{$-$} & \multicolumn{2}{c}{$-$} 
& ~~ $ 6.58 ^{+0.14}_{-0.13}$ 
\vspace*{3pt}  \\
  Mrk 509  & $-$ & \multicolumn{2}{c}{$-$} & \multicolumn{2}{c}{$-$} 
& ~~ $ 7.96 ^{+0.05}_{-0.06}$ 
\vspace*{3pt}  \\
  Mrk 590  & 4.7 & $ 189 $ & $  6 $ & $  7904 $ & $  7  $ 
& ~~ $ 7.14 ^{+0.10}_{-0.09}$ 
\vspace*{3pt}  \\
  Mrk 817  & 4.7 & $ 120 $ & $ 15 $ & $  9409 $ & $ 11  $ 
& ~~ $ 7.55 ^{+0.11}_{-0.12}$ 
\vspace*{3pt} \\
  Akn 120  & 3.0 & $ 221 $ & $ 17 $ & $  9575 $ & $ 15  $ 
& ~~ $ 8.27 ^{+0.08}_{-0.12}$ 
\vspace*{3pt}  \\
  3C 390.3 & 3.6 & $ 273 $ & $ 16 $ & $ 16640 $ & $ 15  $ 
& ~~ $ 8.57^{+0.12}_{-0.21}$ 
\vspace*{3pt} \\

\vspace*{-5pt}
\extline
\hline
\end{tabular}

\end{centering}

{\smallskip \footnotesize $^\ast$ The black hole masses are those
using the r.m.s. BLR widths.  Sources are \citet{onken} for NGC 3227,
NGC 3516 and NGC 4593, \citet{ho99} for NGC 7469 and \citet{kaspi2000}
for the rest.}

\end{table}

\begin{figure*}

\begin{centering}

\rotatebox{90}{\scalebox{0.62}{\includegraphics{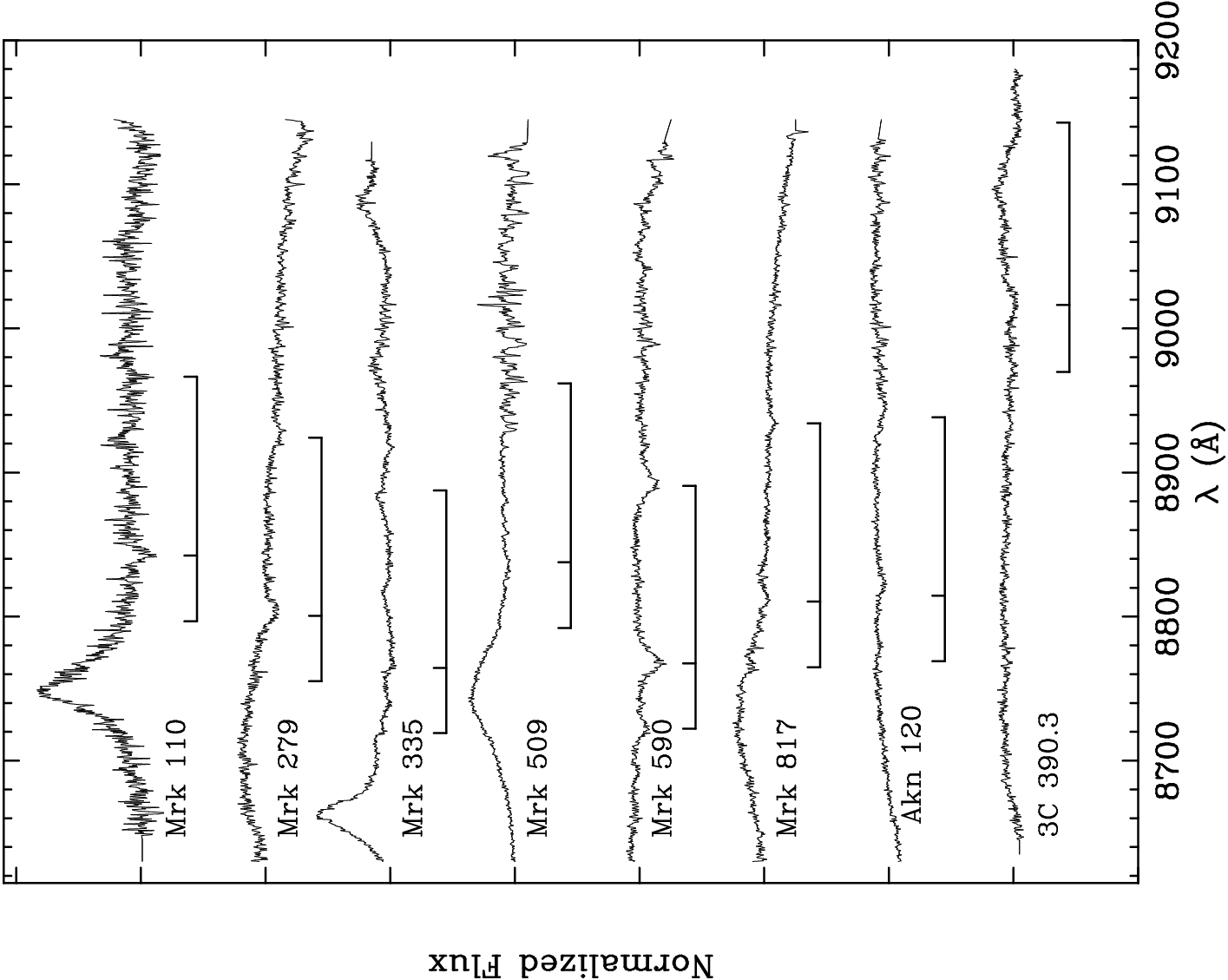}}}

\vspace*{30pt}

\rotatebox{90}{\scalebox{0.62}{\includegraphics{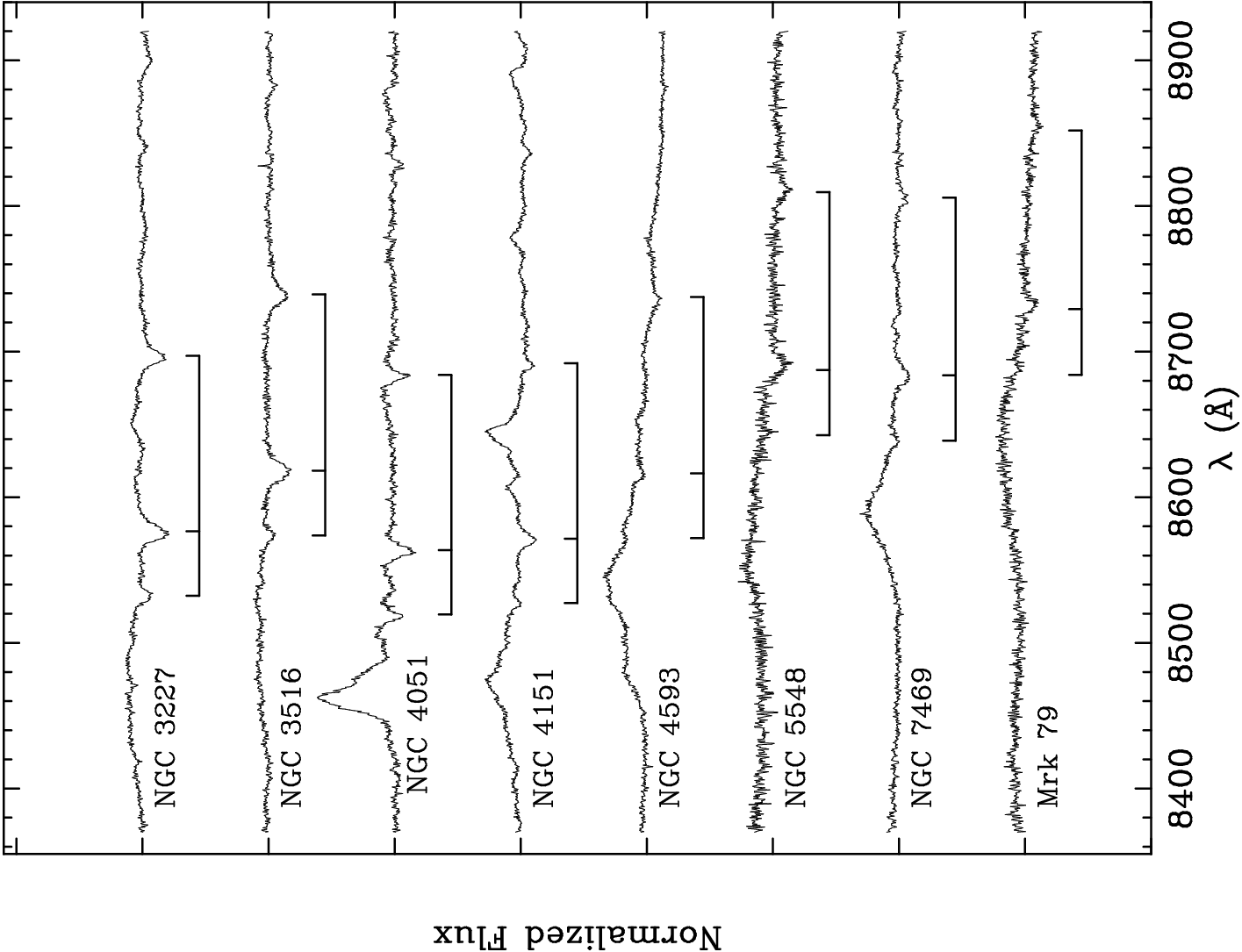}}}

\figcaption{Reduced spectra are presented with telluric
absorption correction and relative flux calibration applied.  The position of
the \Cat \ lines are indicated by the bar with three ticks below each
spectrum. The tick marks on the left hand vertical axis indicate the
zero points for each spectrum, giving a sense of the broad range in
\Cat \ equivalent width.
\label{fig:rawspec}}
\end{centering}

\end{figure*}

\section{Spectroscopic Data}
\label{sec:obs}

\begin{figure*}

\begin{centering}

\rotatebox{90}{\scalebox{0.62}{\includegraphics{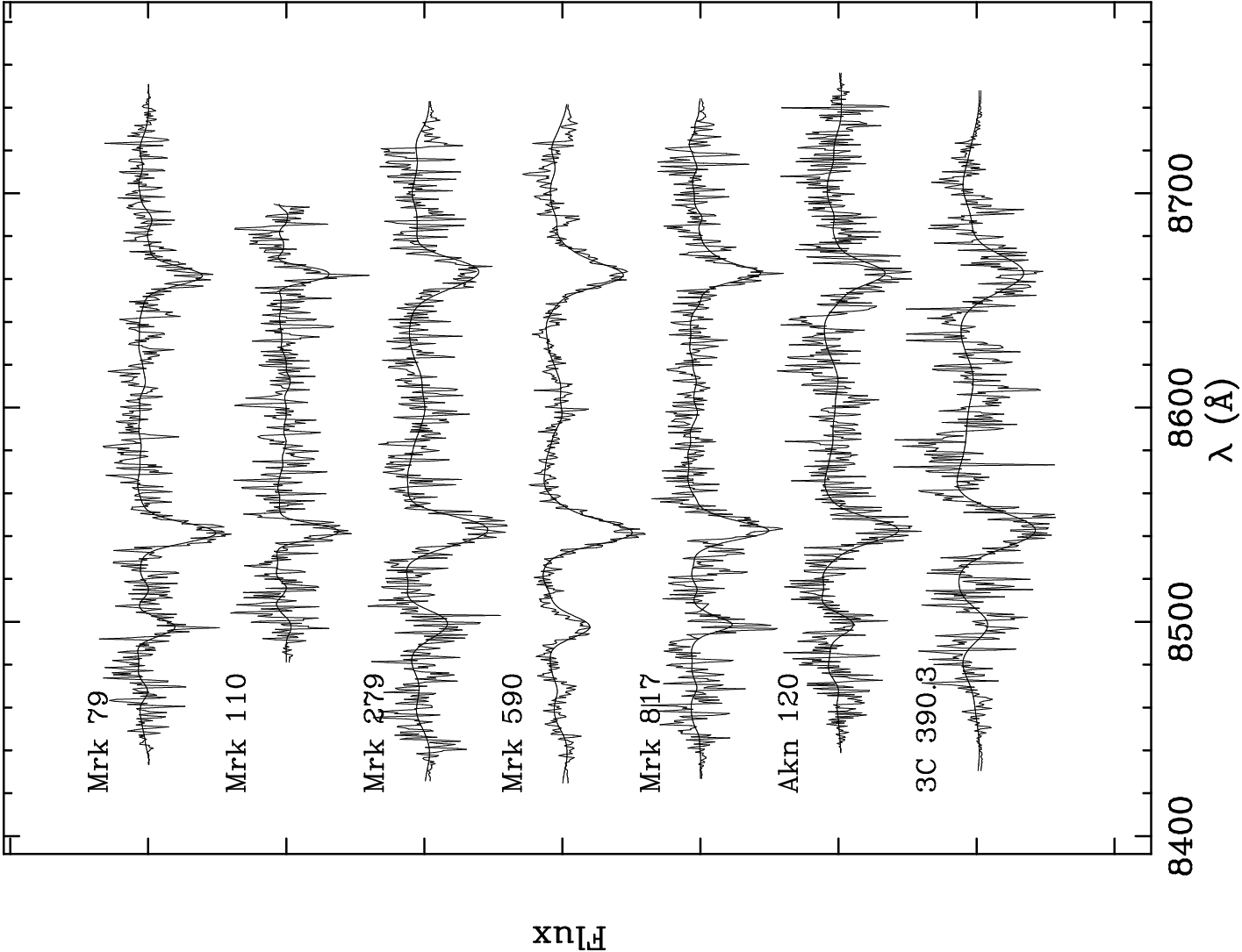}}}

\vspace*{30pt}

\rotatebox{90}{\scalebox{0.62}{\includegraphics{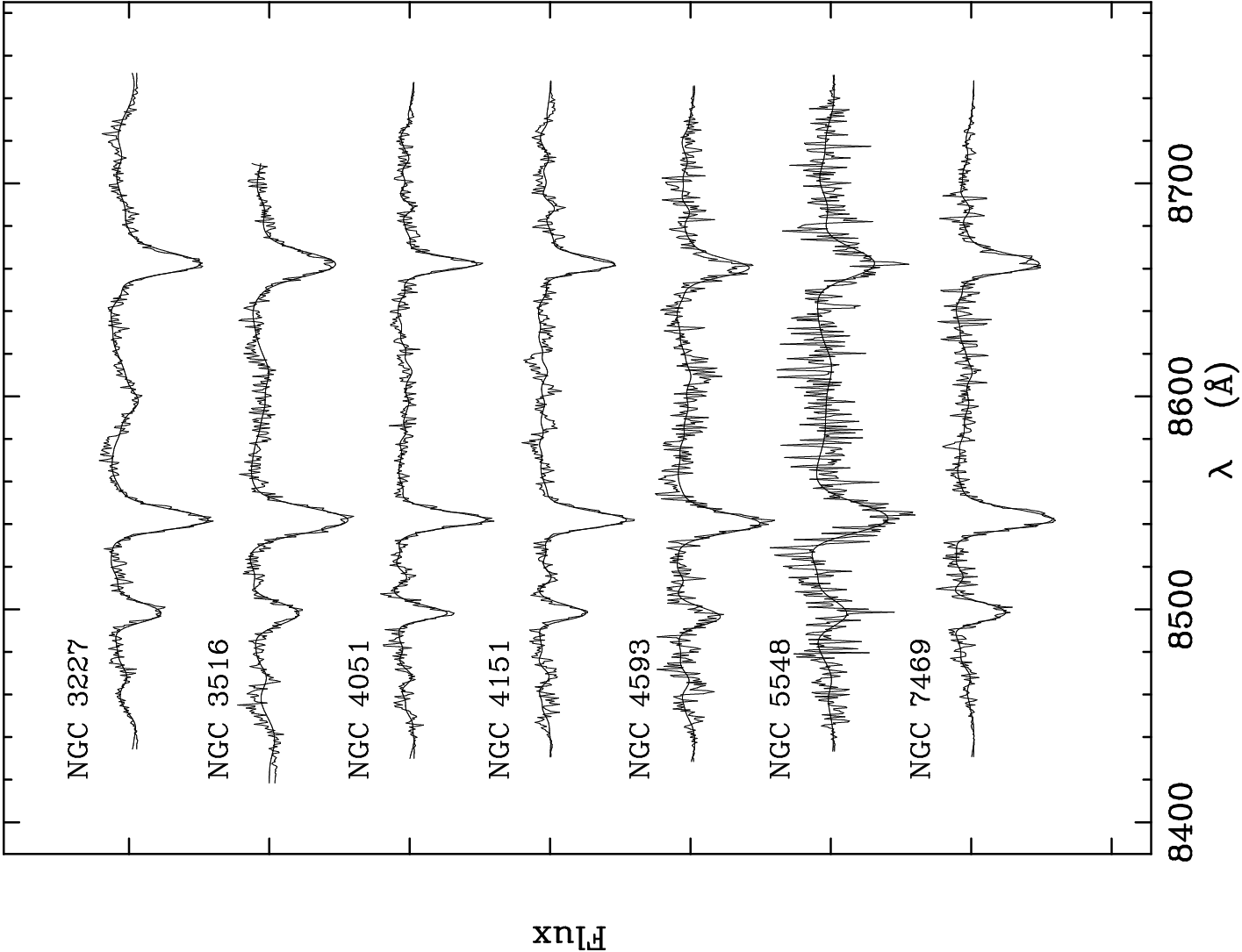}}}

\figcaption{We show the finished spectra which were input to FCQ and
also the broadened stellar template. The spectra have been scaled so
that the middle line in the \Cat \ has approximately the same depth in
each galaxy.
\label{fig:fcqspec}}

\end{centering}
\end{figure*}

\subsection{Observations and Reductions}
\label{sec:reduce}

Longslit spectroscopy of 16 Seyfert 1 galaxies and several stellar
template stars was obtained on two observing runs (U.T. dates April
10-14, 2001 and November 1-4, 2001) using the 4-meter telescope at
Kitt Peak National Observatory.  Our sample is composed of Seyfert 1
galaxies included in long term emission line and continuum monitoring
campaigns for which reverberation mapping estimates of \Mbh\ have been
published \citep{kaspi2000, wpm99, ho99}.  The RC spectrograph was
used with the LB1A CCD which has high quantum efficiency in the
far-red ($\sim 90$\% at 8500 \AA), excellent cosmetics and no fringing
\citep{dey} --- an ideal detector for observing the \Cat \ lines
($\lambda 8498.0, \lambda 8542.1$ and $\lambda 8662.1$ \AA). These
properties allowed us to use a high dispersion grating (BL 380 giving
0.46 \AA/pixel in the raw data) and a narrow slit (1\asec) to achieve
high spectral resolution ($\sim 1.1$\AA \ or $\sim 40$ \kms \ FWHM).
We used an order-blocking filter (OG570) to prevent contamination by
second-order blue light, which will of course be strong in Seyfert 1
spectra.  Exposure times ranged between 3600 seconds for nearby bright
galaxies (e.g. NGC 4051, NGC 3516) and 10,800 sec for fainter objects
(e.g. Mrk 110).  Standard spectroscopic reductions --- bias
subtraction, flat field correction, geometric distortion and
wavelength calibration --- were carried out using IRAF. Conditions
were generally not photometric so only a relative flux calibration was
applied.

The strong telluric emission and absorption features in the far-red
present a difficult reduction problem, particularly for the higher
redshift galaxies in our sample.  For targets with $z \geq 0.016$,
where the \Cat \ and night sky emission lines begin to overlap (except
3C 390.3, see section \ref{sec:galnotes}), sky-subtraction was
accomplished using the ``nod-and-shuffle'' observation mode, developed
by \citet{nodshuffle} and recently implemented at Kitt Peak. Briefly,
by repeatedly re-positioning the telescope (nodding) and shifting the
charge on the detector (shuffling), both the target and the sky are
observed at the same spot along the slit and in the same pixels on the
detector, permitting excellent sky subtraction.  The drawbacks for the
nod-and-shuffle technique are a factor of $\sqrt 2$ increase in the
sky noise and extra overhead time for the telescope nods.  Although
there were difficulties in the implementation of this mode at the
time, we were quite pleased with the results and we recommend use of
the perfected observing mode where sky subtraction is critical and
signal-to-noise is not of primary concern.

Atmospheric absorption bands begin to be a problem at wavelengths
above 8900 \AA, again affecting the higher redshift members of the
sample.  Even if the \Cat \ lines are not directly affected,
correction for these can be important for achieving a good fit to the
continuum on the red side.  We therefore obtained spectra of white
dwarf spectroscopic standard stars, matched in airmass and observation
time to our Seyfert exposures, to use as atmospheric templates.
Experimentation showed that scaling the equivalent width of the
atmospheric bands, by adding a constant value to the template and
renormalizing, improved the results in some cases, helping to recover
good line profiles for the reddest \Cat \ line.  In the case of 3C
390.3, the atmospheric absorption correction was quite critical since
all of the \Cat \ lines lie in a region of strong atmospheric
absorption. In part due to the broad \Cat \ features in this object,
we were able remove the telluric lines successfully.

Reduced, sky-subtracted nuclear spectra corrected for telluric
absorption are shown in figure \ref{fig:rawspec}. The spectra have
been normalized to the mean counts in the spectral region near the
\Cat \ lines, whose positions are indicated for each object.  It is
worth emphasizing that a variety of spectral characteristics are
present in these galaxies. The \Cat \ lines show a range of equivalent
widths presenting a challenge for analysis using the FCQ technique in
some cases.  Also a number of nuclear emission features, including
broad OI $\lambda 8446$, [Fe II] $\lambda 8616$, Paschen and \Cat \
itself in emission, appear in our spectra and must be removed before
analysis with FCQ.  In fact for two galaxies, Mrk 335 and Mrk 509, no
velocity information could be obtained (see section
\ref{sec:galnotes}).  Given the association of AGN and starbursts we
must also keep in mind that the strength of the \Cat \ lines may be
enhanced by young stellar populations rich in late-type supergiant
stars \citep[e.g.][]{tdt,nw04}.

\subsection{Removal of Nuclear Emission}
\label{sec:remnuc}

Even in the far-red, the strength of the nuclear emission from type 1
AGN relative to the contribution from the host galaxy makes obtaining
high signal-to-noise spectra of the stellar absorption features a
difficult task --- effectively, the nuclear continuum is a source of
noise. Also broad and narrow line nuclear emission, even when not
affecting the \Cat \ lines must be carefully removed before the
velocity dispersion measurements can be made.  In some cases, where
the stellar absorption lines are strong and the nuclear emission is
weak, it was possible to include the nucleus of the galaxy in our
longslit extractions without further processing.  For other galaxies,
however, the central rows of the longslit spectra had to be corrected
for contaminating emission features, or when the absorption features
were entirely lost in the noise, excluded altogether.  OI $\lambda
8446$ and [Fe II] $\lambda 8616$ could often be removed by subtraction
of a single or double Gaussian fit.  In other galaxies the blending of
emission lines and continuum was more complex, producing an irregular
continuum shape near the \Cat \ lines. In several cases this required
subtraction of a model of the unresolved nuclear spectrum prior to the
dispersion measurements.

To produce the model, a narrow spectrum, usually two or three pixels
wide centered on the nucleus was extracted. Next, the weak \Cat \
absorption features were removed from the model by subtracting the
spectrum of a stellar template, which was broadened and scaled until
the absorption lines vanished.  Then, using spectra of nearby ninth or
tenth magnitude SAO stars obtained prior to each galaxy exposure, we
produced a seeing profile along the slit.  This was combined with the
model nuclear spectrum and subtracted from the original longslit data
to correct for the contamination by the nuclear emission.  The most
important effect of this procedure is that it removes the irregular
continuum shape from spectra immediately adjacent to the nucleus. This
was particularly helpful in NGC 4051, NGC 4151 and NGC 4593, where the
continuum on the blue side of the spectrum is distinctly uneven
(compare figures \ref{fig:rawspec} and \ref{fig:fcqspec}).  This
allowed us to extract spectra for the velocity dispersion analysis
much closer to the nucleus than would have been possible otherwise,
resulting in higher signal-to-noise and ensuring that as much of the
bulge light was included as possible.

\subsection{Velocity Dispersion Measurements}
\label{ref:fcq}

Velocity dispersions were obtained by analysis of the bulge
line-of-sight velocity distribution (LOSVD) using the
Fourier-Correlation Quotient method \citep{fcq}.  For each galaxy a
spectral region, including only the \Cat \ lines and a small section
of continuum on either side, was used in the analysis (rest wavelength
8460~\AA \ $-$ 8716~\AA). In most cases, the width of the spectrum
along the slit was chosen to cover at least 1 kpc across each
galaxy. For a few nearby galaxies we found that extending the slit all
the way out to $r_e$ (see table \ref{tab:bulgetab}) actually degraded
the signal-to-noise while yielding effectively the same dispersion. In
these cases we opted for the shorter slit length.  For the fainter
targets an optimal extraction maximizing signal-to-noise was used.
Although this does not generally correspond to $r_e$ as used by
\citet{geb2000a} in defining the \Mbserel, given the difficulties
posed by the nuclear emission, it was a reasonable and unavoidable
compromise.  Nevertheless, our spectra include bulge starlight far
enough from the nucleus to be beyond the influence of the black hole,
while excluding emission from the kinematically colder galaxy disk.

Gauss-Hermite polynomials were fitted to the LOSVD to determine the
bulge systemic velocity, \Vs, and the bulge stellar velocity
dispersion, \sigs.  The uncertainties in \Vs \ and \sigs \ were
estimated by running the FCQ algorithm on Monte Carlo simulations of
the data, created using a stellar template convolved with the LOSVD and
adding in noise to match that of the observed spectrum.  The results
for the sample are given in table \ref{tab:s1sigma}, including
\Mbh \ values. Figure \ref{fig:fcqspec} shows
the data used in the FCQ analysis shifted to the rest frame along with
the template broadened by the derived LOSVD.

\begin{figure}
\scalebox{0.45}{\rotatebox{-90}{\includegraphics{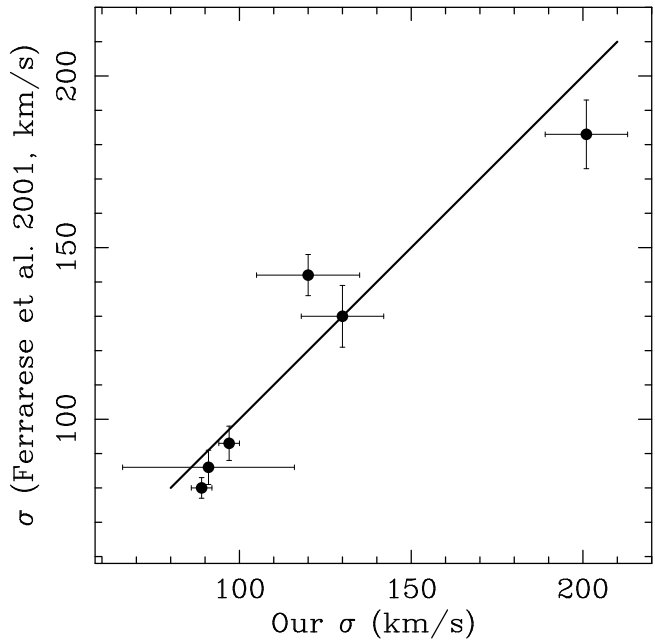}}}
\figcaption{Our \sigs \ measurements are compared with those of 
\citet{fer2001}. The solid line is $X=Y$.
\label{fig:usthem}}
\end{figure}

Velocity dispersions for six of our sample galaxies were published by
\citet{fer2001}.  A comparison of our results with theirs is shown in
figure \ref{fig:usthem}.  We find that our values agree with theirs
quite closely, with a marginal tendency for our dispersions to be
larger at the low \sigs \ end. This may be a consequence of our narrower
slit (they used 2\asec) which excludes more disk light --- a result
that may be important for systems with smaller bulges.

\section{Bulge Magnitudes}
\label{sec:bulmag}

To investigate the reported deviations of AGN from the \Mbh \ -- \Mb \
relation discussed in section \ref{sec:intro}, we have used archival
HST images to measure bulge magnitudes. Ten of the Seyfert 1 galaxies
in our sample were observed in the program of WFPC-2 snapshots
\citep{malkan98} using the F606W filter and another, NGC 4151, was
observed with the F547M filter. The main difficulty in determining the
bulge magnitudes from these images was removing the point source
nucleus, which is unfortunately saturated in the F606W images.  To
accomplish this, model PSFs, created using the Tiny Tim software
package \citep{tinytim}, were shifted, scaled and subtracted from the
images.  The scale factor which produced a minimum r.m.s. deviation in
the residual image in an annulus incorporating the innermost
unsaturated pixels, was chosen as an initial guess.  Since this tends
to overestimate the nuclear luminosity, further trial-and-error
scaling of the PSF around this value allowed us to fine tune the
results; the final best subtraction was judged by examining the
removal of peculiar features of the outer PSF (rays and ring like
structures) in the residual images.  Typically we found that a range
of scale factors produced acceptable results.  The resulting
uncertainty in the bulge magnitudes, which given the
subjective nature of the PSF scaling we uniformly take to be $\pm$0\magx2,
was folded into our error estimates.  Because
the galaxy bulges are reasonably large in the images ($r_e \sim 50 -
100$ pixels in most cases) our results do not depend critically on PSF
subtraction.  The purpose is mainly to remove the extended wings of
the unresolved nuclear emission and recover as much of the inner bulge
light as possible. For most objects, we were able to extend our bulge
profile analysis down to about half an arcsecond.

Other complications included removing the cosmic ray hits from the
images and accounting for emission lines from the Narrow Line Region
(NLR). About 90\% of the cosmic rays were removed automatically using
a script written in IRAF. Of the remaining cosmic rays those close to
the nucleus were removed by hand.  The broad bandpass of the F606W
filter includes the [OIII] $\lambda 5007$, [NII] $\lambda 6548$,
$\lambda 6584$, and H$\alpha$ emission lines which arise in the NLR
and will not be removed with the point-like nucleus.  Using the STSDAS
task CALCPHOT, and the model Seyfert 2 galaxy spectrum provided by the
SYNPHOT package, we were able to estimate the NLR flux by scaling the
model spectrum to match the [OIII] flux tabulated in \citet{w92a}. The
[OIII] fluxes are almost all rated quality ``a'' by Whittle indicating
that they were obtained through large apertures and that there is good
agreement when more than one measurement has been published.  Thus we
are confident that we have accounted for the bulk of the NLR emission.
This provided a correction to our bulge magnitude estimates and is
listed in Table \ref{tab:bulgetab}.

\begin{figure*}
\includegraphics{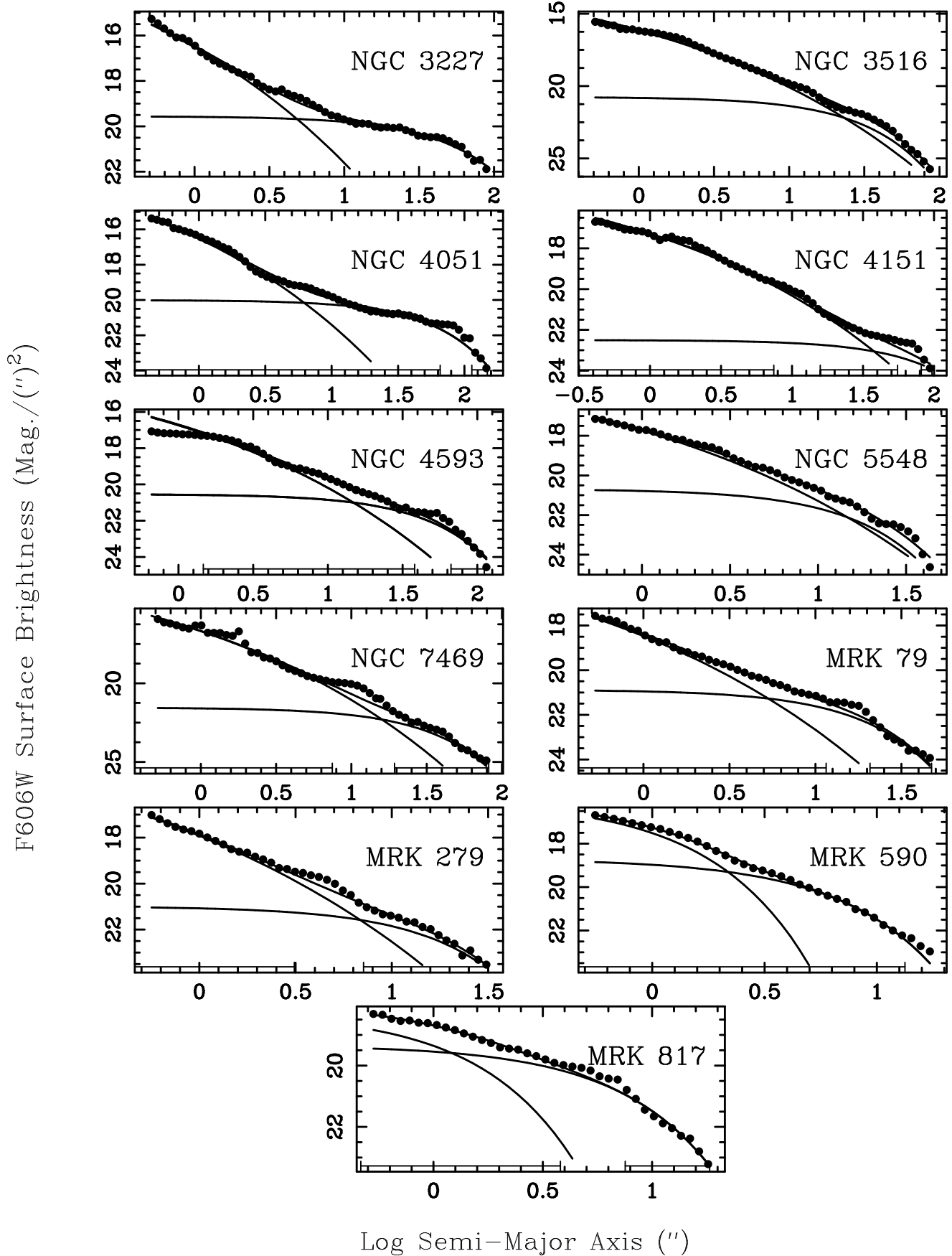} \figcaption{Luminosity profiles and
bulge-disk decomposition are shown for Seyfert 1 galaxies from HST
images.  The emission from the nucleus has already been subtracted
from the profiles. Note the exponential bulges in Mrk 590 and Mrk
817. A horizontal line shows regions included in the fitting, (if not
present all of the data was used) gaps are usually associated with
bars.
\label{fig:lumprof}}
\end{figure*}

Ideally, two-dimensional direct fitting techniques, such as that of
\citet{peng02}, and a less restricted model, e.g. a Sersic function,
could and should be used to investigate the bulge morphologies of
these galaxies.  We note in the descriptions of the individual
galaxies below (section \ref{sec:galnotes}) that the bulge profiles of
Mrk 590 and Mrk 817 are better fit by exponentials than $r^{1/4}$
laws. Also some display inner spirals, dust lanes, lenses and bars that further
complicate the analysis.  However, accounting for these difficulties
and the uncertainties associated with NLR emission and saturated PSFs
is outside the scope of our program. Thus, in this paper, we adopt
either the $r^{1/4}$ law or exponential fits to the azimuthally
averaged luminosity profiles and use these to estimate bulge
magnitude.

We used the IRAF task ELLIPSE to produce luminosity profiles from the
image isophotes which were then fitted with the sum of an exponential
or $r^{1/4}$-law bulge and an exponential disk.  In carrying out the
fits it is important to avoid getting trapped in spurious minima of
$\chi^2$.  The procedure we adopted to ensure meaningful results, was
to fit the disk region first. Then holding the disk parameters fixed
we included a fit to the bulge.  As a final step, we fitted both bulge
and disk simultaneously allowing all parameters to vary, using the
parameter values from the previous steps as the initial guess.  The
luminosity profiles and fits are plotted in figure \ref{fig:lumprof}.
Note the presence of bars or inner ovals in a few galaxies (e.g. NGC
4593, Mrk 79, NGC 7469), seen as a run of excess light over the fit,
between the bulge and disk.  We generally excluded these regions from
the analysis.  Bulge magnitudes were then estimated from the scale
length and surface brightness parameters of the fits, assuming $H_0 =
75$~km~s$^{-1}$~Mpc$^{-1}$ \citep[as used by]
[to calculate black hole mass]{kaspi2000}. Foreground extinction from
the Milky Way has been removed, but no correction for dust within the
host galaxy has been applied. The results are given in table
\ref{tab:bulgetab} and notes on the characteristics of individual
galaxies are given in section \ref{sec:galnotes}.

We estimated the error in the bulge magnitudes using Monte Carlo
simulations.  Random noise with the same r.m.s. deviation as the fit
to the luminosity profiles was added to the data and the profile was
refitted. The resulting standard deviations in the fit parameters
after 30 trials was then propagated through to estimate the error in
bulge magnitudes from the fitting. These were then added in quadrature
to the uncertainty in the nuclear magnitudes and adopted as the
uncertainties listed in table \ref{tab:bulgetab}. We also
include an uncertainty of 15\% in the distances of each galaxy
to compute the final errors in the absolute magnitudes.

\begin{table}

\caption{\label{tab:bulgetab}}

\begin{centering}

\begin{tabular}{lccccc}
\multicolumn{6}{c}{Bulge Magnitudes in Seyfert 1 Galaxies}
\smallskip \\
\hline 
\extline
\null Name & $\mu_e^{\rm a}$ or $\mu_0$ & $r_e^{\rm b}$ or $r_0$ & $m_{bul}$ 
& $\Delta m_{NLR}$ & $M_{bul}$ \\
\null & (mag.) & ($\prime \prime$) 
& (mag.) & (mag.) & (mag.) \\
\null & (F606W) & & (F606W) & (F606W) & (F606W) \smallskip
\eol
\hline \hline
\extline
 NGC 3227 &  18.31 &    2.63 &   13.24 &    0.09 &  $-$18.19\\
\smallskip   &   $\pm$0.16 &    $\pm$0.26 &    $\pm$0.36 &   &    $\pm$0.49\\
 NGC 3516 &  19.43 &   7.41 &   11.91 &    0.02 &  $-$21.19\\
\smallskip   &   $\pm$0.12 &   $\pm$0.55 &    $\pm$0.29 &   &    $\pm$0.44\\
 NGC 4051 &  18.58 &   3.10 &   13.08 &    0.05 &  $-$18.53\\
\smallskip   &   $\pm$0.12 &   $\pm$0.23 &    $\pm$0.30 &     &   $\pm$0.44 \\
 NGC 4151 &  21.40 &  20.86 &   11.52 &    0.00 &  $-$20.20\\
\smallskip   &   $\pm$0.09 &   $\pm$1.59 &    $\pm$0.28 &     &    $\pm$0.43\\
 NGC 4593 &  20.63 &  12.44 &   12.01 &    0.01 &  $-$20.87\\
\smallskip   &   $\pm$0.17 &   $\pm$1.30 &    $\pm$0.37 &    &    $\pm$0.46\\ 
NGC 5548 &  21.49 &  11.03 &   12.89 &    0.06 &  $-$21.31\\
\smallskip &   $\pm$0.26 &   $\pm$3.20 &    $\pm$0.71 &     &    $\pm$0.78\\
NGC 7469 &  19.66 &   5.21 &   12.93 &    0.06 &  $-$21.15\\
\smallskip &   $\pm$0.14 &   $\pm$0.44 &    $\pm$0.32 &     &    $\pm$0.45\\
 Mrk 79  &    21.41 & 5.68 &   14.46 &    0.16 &  $-$20.29\\
\smallskip  &    $\pm$0.28 & $\pm$0.70 &    $\pm$0.45 &    &  $\pm$0.56 \\
 Mrk 279 &  20.31 &   3.89 &   14.53 &    0.06 &  $-$20.86\\
\smallskip  &   $\pm$0.15 &   $\pm$0.41 &    $\pm$0.39 &     &    $\pm$0.51\\
 Mrk 590$^\ast$ &    15.99 &  0.71&  14.79 & 0.03 &   $-$20.25 \\
\smallskip  &     $\pm$0.03 & $\pm$0.02 & $\pm$0.21 & & $\pm$0.39 \\
 Mrk 817$^\ast$ &    18.25 & 0.98 &   16.54 & 0.44 &   $-$18.58 \\
\smallskip & $\pm$0.09 & $\pm$0.13 & $\pm$0.39 & & $\pm$0.51 \\
\extline 
\hline 
\multicolumn{6}{l}{$\rm ^a$}{$\mu_e$ and $\mu_0$ are the surface brightness
fitting parameters for r$^{1/4}$-law} \\
\multicolumn{6}{l}{and exponential forms of the
luminosity profile respectively.} \\
\multicolumn{6}{l}{$\rm ^b$}{$r_e$ and $r_0$ are the scale length 
fitting parameters  for r$^{1/4}$-law and } \\ 
\multicolumn{6}{l}{exponential forms of the
luminosity profile respectively. } \\
\multicolumn{6}{l}{$^\ast$ indicates an exponential bulge was fitted to the
data; all others } \\
\multicolumn{6}{l}{are $r^{1/4}$ law fits.} \\
\end{tabular}

\end{centering}

\end{table}

Bulge magnitudes for several of the same Seyfert galaxies were
tabulated by \citet{mcluredunlop02}, some measured using the same HST
images, while others were taken from the photographic study of
\citet{bagbagand}.  In figure \ref{fig:usthemmag}, we plot our values
against the \citet{mcluredunlop02} results, where we have corrected
their values to our choice of $H_0$ and to the F606W bandpass (since
they quote \Mb \ in the $R$ bandpass). The solid line shows $X=Y$, and
suggests no systematic trends, although the large scatter (r.m.s. $ =
$1\magx 3, compared with the combined error estimate of 1\magx 0) does
indicate some disagreement in a number of galaxies.  Although
\citet{mcluredunlop02} were careful to avoid saturating the nucleus in
their own observations of quasar hosts, they do not discuss how they
handled the saturated nuclei in the \citep{malkan98} snapshots.  Also,
in Mrk 817 we have a rather large correction for NLR emission, which
to our knowledge was not considered by \citet{mcluredunlop02}.  In any
case the large scatter, while disappointing, is not far outside what
might be expected from the uncertainties in the measurements.

\section{Notes on Individual Seyfert 1 Galaxies}
\label{sec:galnotes}

{\it NGC 3227} --- The active continuum is weak, and the Ca II triplet
lines are seen at nearly full strength \citep{nw04}. Note the presence of
Paschen absorption lines in figure \ref{fig:fcqspec} (particularly P14
$\lambda 8598.4$ between the two stronger \Cat \ lines) indicating the
presence of a young stellar population. An F5 III star was used for a
velocity template in this case. The bulge of this galaxy is
surprisingly small and faint given its Hubble type 
\citep[SABa][]{rc3} and \sigs, perhaps as a result of the dusty
nuclear regions \citet{malkan98}.

{\it NGC 4051} --- Weak Ca II triplet in emission can be seen in this
object as well as strong and relatively narrow OI $\lambda $ 8446 and
[Fe II] $\lambda $ 8618 emission. All of these were removed by
combining model nuclear emission with a seeing star as described in
section \ref{sec:remnuc}.

\begin{figure}
\scalebox{0.45}{\rotatebox{-90}{\includegraphics{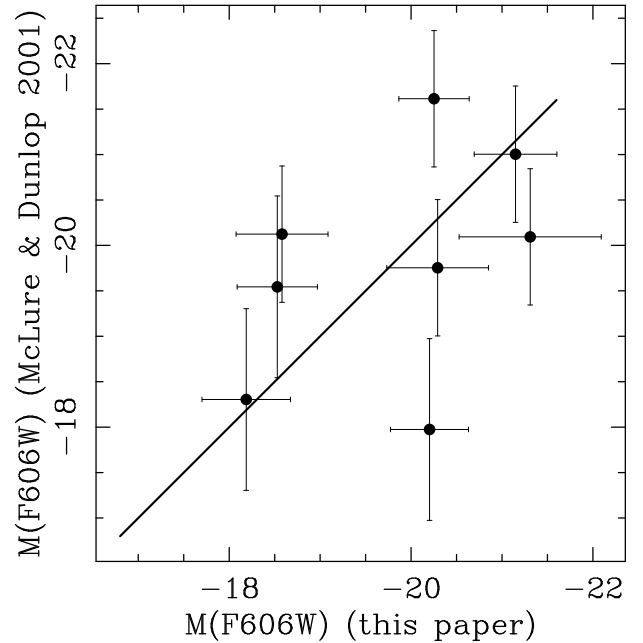}}}
\figcaption{Our \Mb \ measurements are compared with those of 
\citet{mcluredunlop02} rescaled to rescaled to
$H_0=75 \, \rm km \, s^{-1} Mpc^{-1}$. The solid line is $X=Y$.
\label{fig:usthemmag}}
\end{figure}

{\it NGC 4151} --- Nuclear emission features in the \Cat \ spectrum
were removed by combining model nuclear emission with a seeing
star. Long and short exposure images for this object were obtained with
the F547M filter. This allowed an accurate subtraction of the
nucleus. Also this filter excludes most of the optical emission lines
so no correction was applied for the NLR. The magnitude in table
\ref{tab:bulgetab} has been converted to the F606W bandpass using
CALCPHOT assuming the spectrum is well represented by the Sb template
in the SYNPHOT package. This is the most discrepant point in figure
\ref{fig:usthemmag}, for which \citet{mcluredunlop02} used the
bulge-disk decomposition of \citet{bagbagand} based on ground
based photographic imaging.

{\it NGC 4593} --- Nuclear emission features were removed by
combining model nuclear emission with a seeing star. 

{\it NGC 5548} --- Since the velocity dispersion falls off quite
rapidly with distance from the nucleus a narrow spectral extraction
was used.  

{\it NGC 7469} --- The position angle and longslit extraction were chosen
to avoid knots in the well-known starburst ring \citep{7469ring}. OI
$\lambda 8446$ was removed with Gaussian fit.

{\it Mrk 110} --- This was the poorest spectrum in our sample and as a
consequence we quote a large error estimate for the velocity
dispersion. OI $\lambda 8446$ removed with Gaussian fit.

{\it Mrk 279} --- Broad, weak OI $\lambda 8446$ was removed in normal 
continuum subtraction. 

{\it Mrk 335} --- \Cat \ in emission \citep{malkanfilip83} could not be removed.

{\it Mrk 509} --- No \Cat \ features were detected due to the strong 
nuclear continuum.

{\it Mrk 590} --- The bulge of this galaxy is better fit by a
exponential than an $r^{1/4}$ law.

{\it Mrk 817} --- OI $\lambda 8446$ was removed with double Gaussian
fit. The bulge is better fit by a exponential than an $r^{1/4}$ law.

{\it 3C 390.3} --- Due to its high redshift, the \Cat \ lines fall in
a region relatively free of night sky emission lines eliminating the
need for nod-and-shuffle sky subtraction.  However, the atmospheric
absorption is severe; careful division by atmospheric
template was remarkably successful.

\section{The \Mbh \ -- \sigs\ Relation for Active Galaxies}
\label{sec:mbhsig}

\begin{figure}
\scalebox{0.45}{\includegraphics{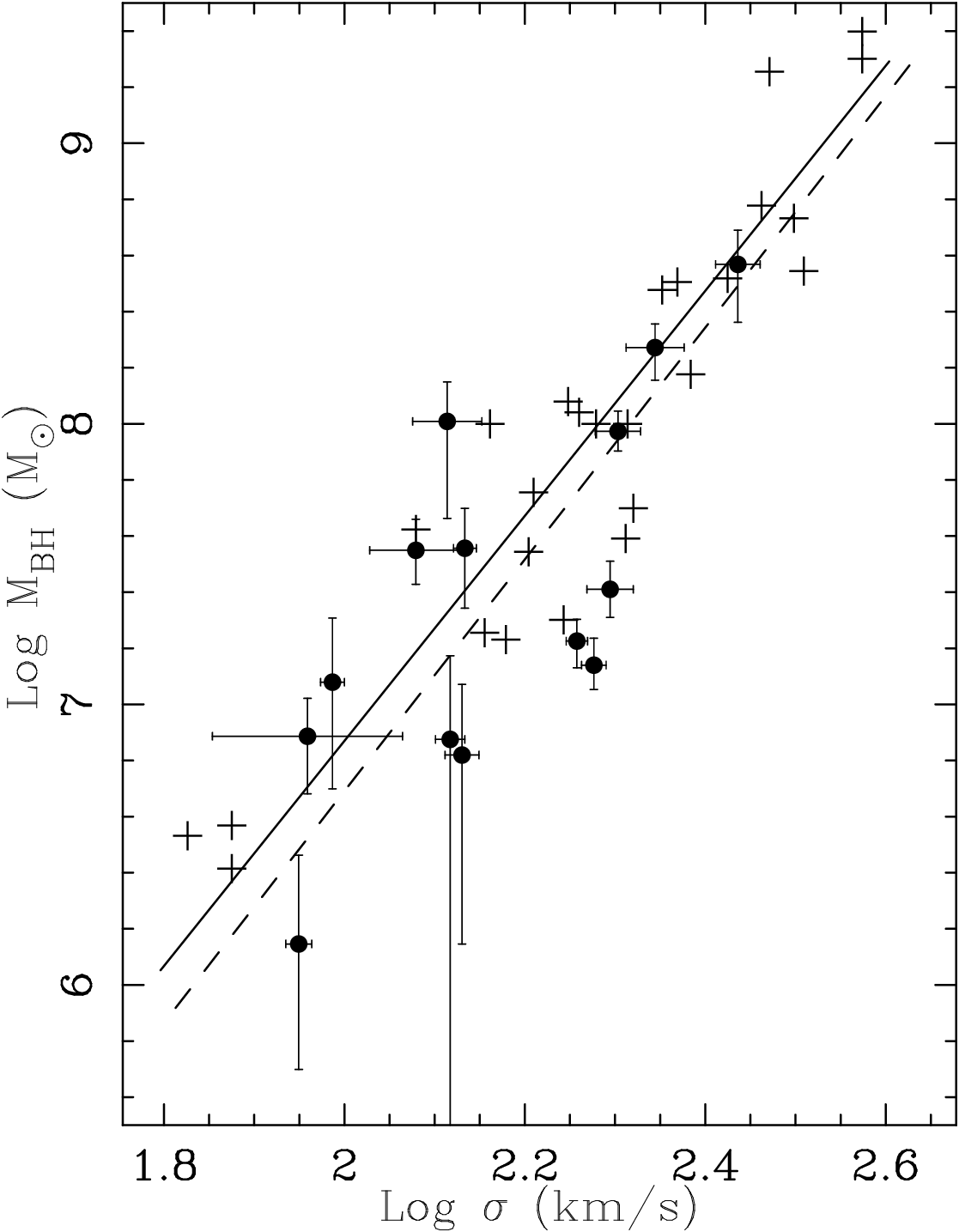}} 
\figcaption{Black hole
mass is plotted against velocity dispersion for Seyfert galaxies from
Table \ref{tab:s1sigma}, shown as $\bullet$ symbols and for normal 
elliptical and S0 galaxies from \citet{geb2000a}, shown as
$+$ symbols.  The black hole masses for Seyfert galaxies
are determined from reverberation mapping, while the \citet{geb2000a}
masses are from dynamical modeling of the nuclear stellar kinematics.
The solid line is the relation calculated by \citet{tremaine} and the 
dashed line is a fit to the Seyferts using the ordinary least-squares
bisector method \citep{isobe90}.
\label{fig:mbhsigagn}} 
\end{figure}

In Figure \ref{fig:mbhsigagn} we plot as $\bullet$ symbols our \sigs\
values versus \Mbh\ values from reverberation mapping. We use the \Mbh
\ values directly from the literature, but note that these are
calculated assuming random orbits for BLR clouds following equation 5
in \citet{kaspi2000} \citep[see also][]{wpm99}.  We discuss the
significance of this assumption in section \ref{sec:mbhmb}.
Furthermore, as mentioned in the note to table \ref{tab:s1sigma}, the
black hole masses are those calculated using the r.m.s. H$\beta$ FWHM
rather than the mean values, i.e. the variable component of the line
\citep[see][]{wpm99}.  Having tried both, we find no significant
differences in our results using one or the other.  For comparison, we
also plot as $+$ symbols, elliptical and S0 galaxies from
\citet{geb2000a}, where \Mbh\ has been determined from stellar
dynamical orbit-based modeling. The solid line is the derivation of
the \Mbserel\ from \citet{tremaine} and the dashed line is a fit to
the Seyferts using the ordinary least squares bisector method
\citep[hereafter OLSB,][]{isobe90}. The fourteen Seyferts, by
themselves, show a strong correlation (linear correlation coefficient,
$R=0.73$ for the r.m.s. masses with probability of no correlation,
$P_{null} = 0.29$\% and slightly better for the mean masses $R=0.80$,
and $P_{null} = 0.065$\%).

In figure \ref{fig:mbhsigagn} there appears to be 
no significant evidence for an offset for Seyferts from the
\citet{tremaine} relation. Computing the residuals of the AGN to the
solid line we find an average shift of $\Delta \log $\Mbh $=0.21$ in
the direction of lower black hole masses at a given \sigs, with a
standard deviation of $0.46$ dex for the 14 AGN.  Clearly, the
significance of such a shift is marginal and distinctly not in
agreement with the factor of 5 found by \citet{ho99} using the \Mbh\
-- \Mb\ relation.

The exact value of the slope of \Mbh\ -- \sigs\ has been a matter of
some debate.  The slope, $4.1\pm0.5$ for the AGN sample is not
significantly different from the slope for the normal galaxies,
although given the uncertainty, it is beyond the ability of this
dataset to provide a decisive result.  A larger sample of AGN might
well provide interesting constraints.

\begin{figure}
\scalebox{0.45}{\includegraphics{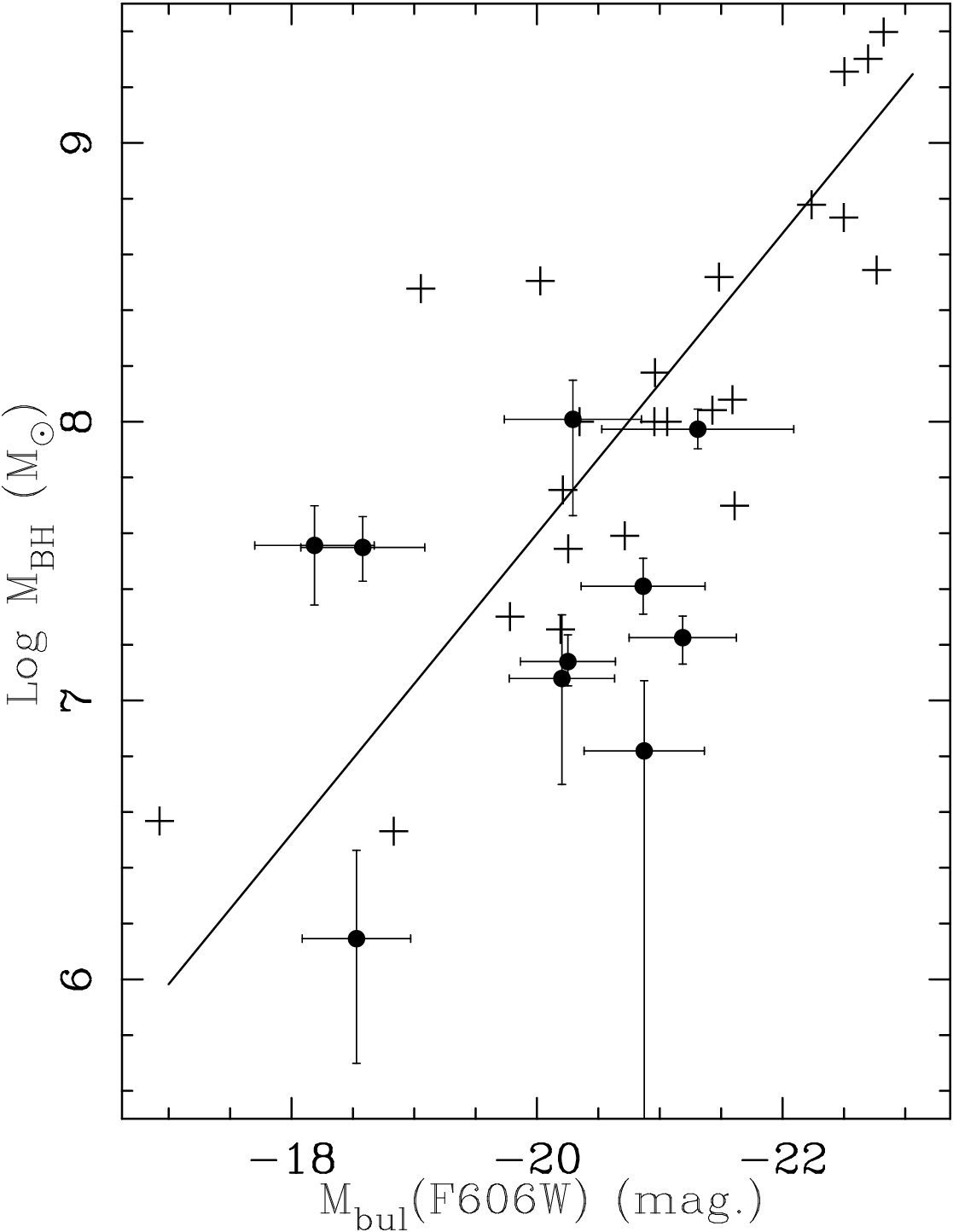}} \figcaption{Black hole
mass is plotted against bulge absolute magnitude for Seyfert galaxies
($\bullet$ symbols) and the \citet{geb2000a} sample ($+$ symbols) rescaled to
$H_0=75 \, \rm km \, s^{-1} Mpc^{-1}$. The solid line is a fit to the 
elliptical galaxies.
\label{fig:mbhmbul}}
\end{figure}

These results effectively confirm the conclusions of \citet{geb2000b},
\citet{nel2000} and \citet{fer2001} that there is no distinct
difference in the \Mbserel \ between active and non-active galaxies.
Furthermore, it suggests that the reverberation mapping technique for
determining black hole masses in AGN produces results which are
consistent with masses determined by stellar and gas dynamical
techniques.  We should emphasize, however, that this is an indirect
check on reverberation mapping.  A direct verification may be possible
should a target of reverberation mapping campaigns go into a low
luminosity phase, allowing application of standard stellar dynamical
analysis. Indeed an HST cycle 12 proposal involving target of
opportunity observations of NGC 3227 and NGC 4151 has been approved
(Prop. No. 9849, PI B. Peterson).

Our results, of course, are very encouraging for investigations of
the role of black hole mass in AGN. However, the sample used here is
by necessity composed of lower luminosity Seyfert 1 galaxies. Further
study of the \Mbserel \ is needed to validate its application
to more luminous AGN.

\section{The \Mbh \ -- \Mb \ Relation for AGN Revisited and 
the Faber-Jackson Relation}
\label{sec:mbhmb}

It is essentially conventional wisdom that active nuclei are
preferentially found in bulge dominated galaxies \citep[e.g.][]{ho5}.
Confirmation of the \Mbssrel \ for Seyferts is important support for this
tenet of AGN physics.  It strongly reinforces the notion that the
bulge gravitational potential plays an important role in determining the
properties of activity in galaxies.  However, if we accept that the
\Mbssrel \ is the same for Seyferts as it is for normal bulges and
ellipticals, what do we make of previous claims of smaller \Mbh \
values in AGN for a given bulge luminosity, discussed in the
introduction to this paper?

\begin{figure}
\scalebox{0.45}{\includegraphics{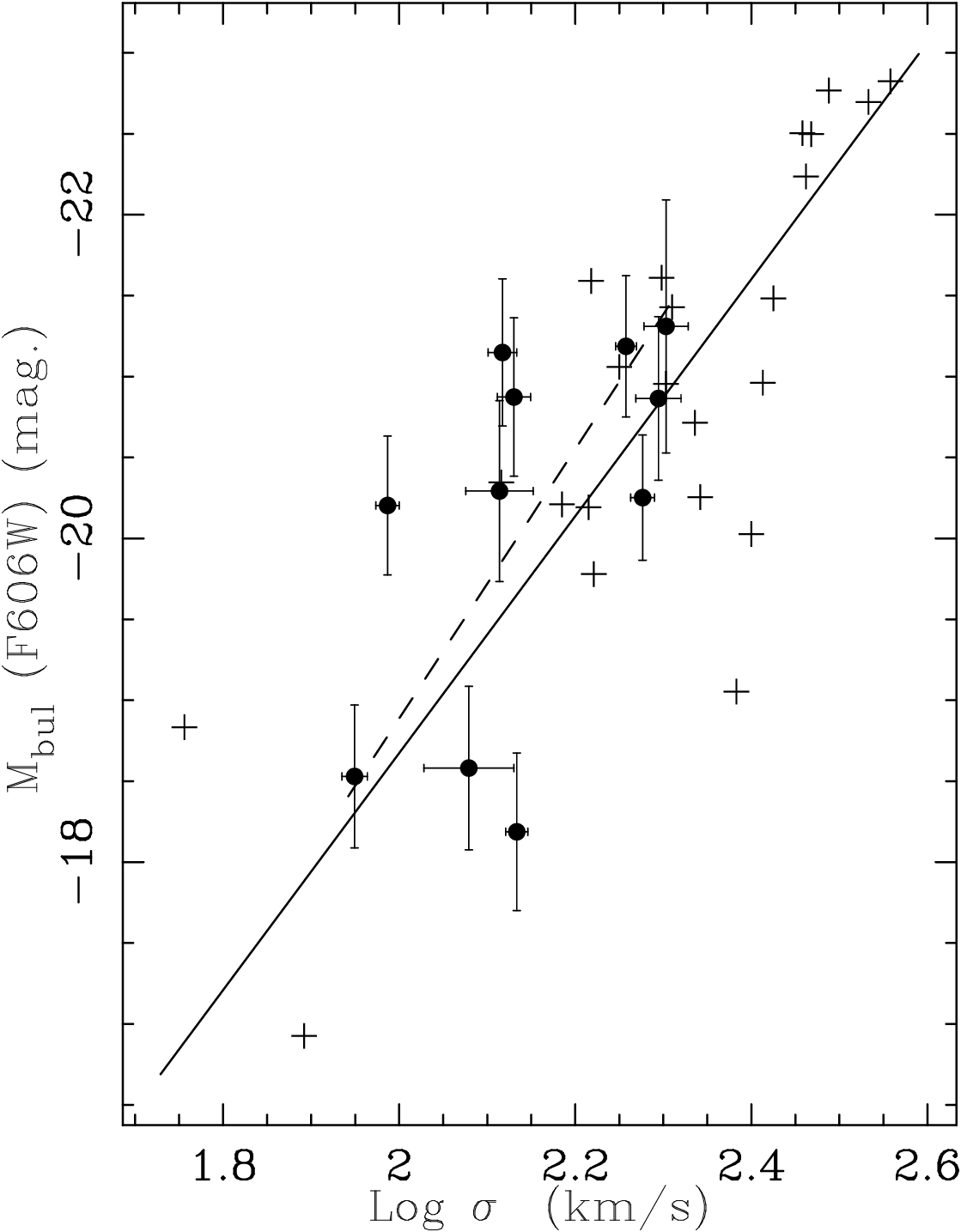}} 
\figcaption{The Faber-Jackson
plot for Seyfert 1 galaxies from Table \ref{tab:bulgetab} ($\bullet$
symbols) and for elliptical and S0 galaxies from \citet[][$+$
symbols]{geb2000a}. The solid line is a fit to the elliptical
sample and the dashed line is a fit to the Seyfert galaxies.
The Seyfert galaxies in the sample show a tendency for
brighter \Mb \ at a given \sigs.
\label{fig:fj}}
\end{figure}

In Figure \ref{fig:mbhmbul} we plot \Mb \ for the Seyfert 1 galaxies
in table \ref{tab:bulgetab}, versus the reverberation mapping \Mbh\
values.  Again for comparison, ellipticals and bulges from the
\citet{geb2000a} sample, are plotted as $+$ symbols,
where we have converted their absolute $V$ magnitudes as published in
\citet{faber97} to the F606W bandpass and scaled to $H_0=75 \, \rm km
\, s^{-1} Mpc^{-1}$. Given the narrow range of colors for the
\citet{geb2000a} sample ($\langle B-V \rangle = 0$\magx$95$,
std. dev. 0\magx047) and the expectation that the Seyfert bulge colors
are likely to be more scattered, we choose to apply the correction to
the ellipticals. The conversion was determined using the template
elliptical galaxy from the SYNPHOT package and the CALCPHOT task
($M_{\rm F606W} = V - $0\magx33).

Although not a strong trend, the majority of Seyferts lie below and to
the right of the OLSB fit to the ellipticals, shown as the solid
line. This is consistent with the report by \citet{ho99} of lower black
hole masses at a given bulge magnitude.  The median deviation in black
hole mass for the Seyferts from the fit to the normal galaxies is
$0.4$ dex, or about a factor of 2.5. There is really no
significant correlation for the Seyferts, (linear correlation
coefficient $R=-0.16$), compared with a strong trend found for the
ellipticals ($R=-0.80$). This may be due in part to the smaller sample
size, but this cannot be the whole story since the Seyferts show a
strong correlation between \Mbh \ and \sigs (see section
\ref{sec:mbhsig}).

We suggest that it is the bulge luminosities, not the black hole
masses, which are the origin of the deviation.  Taken in this way the
bulges of Seyferts are on average 0\magx7 brighter than normal
ellipticals with similar black hole masses based on the distribution
of points in figure \ref{fig:mbhmbul}.  Strong support for this
interpretation is found by plotting the Faber-Jackson diagram, \sigs \
vs. \Mb, for both samples, shown in fig. \ref{fig:fj}. We find that
that at a given \sigs\ the Seyfert galaxies tend to lie above the fit
to the normal galaxies.  The mean residual in \Mb\ for the Seyferts
relative to the normal galaxy fit is $-$0\magx4. Furthermore there is
a moderately strong correlation for the Seyferts, ($R=-0.6$). This
improvement over the results for \Mb \ vs. \Mbh \ naturally suggests
that \Mb \ is more closely related to \sigs \ than to \Mbh, and that
the increase in bulge luminosities with black hole masses is secondary
and follows through their mutual trends with velocity dispersion.

The result is identical to that of \citet{nw96} who found a similar
offset for a much larger sample, consisting of mostly Seyfert 2s. The
result also confirms the earlier conclusions of \citet{w92b,w92c} who
used galaxy rotation amplitude and \fwoiii as virial parameters, to
demonstrate a tendency for brighter bulges in Seyferts than normal
galaxies of the same bulge mass.  Some ambiguity remained, however,
since the bulge magnitudes in those studies were not obtained by
bulge/disk decomposition. Instead, after correcting for nuclear
emission, the relation between bulge-to-disk ratio and Hubble stage of
\citet{simdevauc} was used to estimate bulge magnitudes.  This
technique has been criticized by \citet{mcluredunlop02} and
\citet{wandel02} who suggest that it results in overestimates of
\Mb. The present confirmation of the offset in the Faber-Jackson
relation for Seyferts using explicit bulge-disk-nucleus separation,
however, suggests that while there is admittedly significant scatter,
any systematic errors are small.

In contrast to the \citet{ho99} result, \citet{mcluredunlop02} and
\citet{wandel02} claim no deviation in the \Mbh \ --- \Mb \ relation
for Seyfert and normal galaxies. We can reconcile this conflict by
recognizing that \citet{mcluredunlop02} choose to calculate \Mbh \
assuming a disk-like morphology and purely rotational kinematics for
the BLR as opposed to random orbits assumed by \citet{kaspi2000}. Thus
the observed line widths depend on the orbital velocity, $V$, and the
inclination of the BLR disk. After making assumptions about the range
and mean of the distribution of inclinations, they adopt a relation
between orbital velocity and the BLR line widths, $V=1.5\times FWHM$,
distinctly different from the case for random cloud orbits,
$V=\sqrt{3}/2 \times FWHM$.  This results in \Mbh \ values that are a
factor of 3 {\it larger} than the standard reverberation mapping
calculations in which the BLR velocity field is assumed to be random.
Therefore, if we were to couple their bulge measurements with the
\Mbh\ values calculated assuming random BLR motions, we should also
get brighter bulges at a given \Mbh \ for AGN.  Of course the specific
form of the BLR velocity field is unknown.  Our results, at least for
the case of random cloud motions, are consistent with an identical
\Mbserel \ for Seyfert and normal galaxies, and a tendency for
brighter bulges in Seyferts.

A different argument applies to the \citet{wandel02} analysis, which
assumes random motions dominate in the BLR. This study reports a
correlation between H$\beta$ line width and the suspected overestimate
of \Mb\ resulting from the application of the \citet{simdevauc}
$B/T$-morphology relation discussed above. This correlation is then
used to correct the \Mb\ values, resulting in a reduction of the bulge
luminosities for much of the sample.  However, the physical link
between BLR kinematics and differences in $B/T$ is not established.
It is therefore unclear why such a correction should be
required. Removing this correction from Wandel's results should also
result in relatively brighter bulges for AGN.
 
The most straightforward interpretation of these results is that
Seyfert galaxies have, on average, brighter bulges as a result of
lower $M/L$ ratios. Thus, the difference is one of stellar population,
indicating that Seyferts are more likely than non-active galaxies to
have had recent episodes of nuclear star formation. This is certainly
plausible, given the voluminous research on the link between AGN and
starburst galaxies.  It is informative to use model stellar
populations to estimate the age and mass fraction of the bulge
required to produce a brightening of this degree. Guy Worthey's
Dial-a-Galaxy website \citep{worthey94} allows users to mix stellar
population models of different ages, metallicities, etc., to estimate
the changes to the integrated light properties of galaxy spectra. A
simple run, with two solar metallicity populations of 1 Gyr and 12
Gyr, shows that if 15\% of the total mass is from the young population
the $R$-band magnitude will be $\sim 0$\magx$5$ brighter than a
population composed of completely old stars. Obviously more complex
and realistic models could produce similar results. Nevertheless, this
simple estimate indicates a typical size and age for a burst of star
formation to produce the shift seen in the Seyfert sample.

It is interesting to note that a number of recent morphological
studies of Seyfert galaxies have found no evidence for morphological
triggers of nuclear starbursts --- galaxy interactions and bars,
including sub-kpc scale bars
\citep[e.g.][]{ho5,poggemartini,reganmulc}.  Although it is
conceivable that there is a time delay between a galaxy interaction
and the onset of nuclear fueling, the result is troubling since we
might expect at least some evidence of the disturbance to remain.  A
possible solution was hinted at in \citet{nw96} based on the fact that
Seyferts are not found outside the general envelope of scatter for
normal galaxies in the Faber-Jackson plot.  Considering the residuals
to the Faber-Jackson relation, they found that the distribution of
$\Delta \log \sigma_*$ is considerably narrower yet overlapping that
for non-active galaxies, a detail which is reproduced in Figure
\ref{fig:fj}.  Assuming that the spheriods of normal galaxies exhibit
a broad range of star-formation histories, perhaps  Seyferts are more
akin to normal spirals with above average central star formation rates
than to massive starbursts initiated by strong interactions or bars.


\section{Summary}
\label{sec:summary}

We have obtained new measurements of the bulge stellar velocity
dispersion in Seyfert 1 galaxies with nuclear black hole masses
determined using the reverberation mapping technique. Our results have
shown that Seyfert galaxies follow the same correlation between \Mbh \
and \sigs \ as for non-active galaxies and that reverberation mapping yields
black hole masses which are consistent with masses determined through
stellar dynamical modeling techniques. We also investigated the
origin of previous assertions that the masses of black holes in active
galaxies were smaller than those in normal galaxies (or at least
underestimated) at a given bulge luminosity. We find that the 
best explanation is that the bulges are brighter on average in Seyfert 
galaxies, suggesting lower $M/L$ ratios,  most likely as a result of 
higher star formation rates in their recent history. 

\acknowledgments

We would like to thank the Kitt Peak staff for their help with the
observations using the new LB1A chip. In particular we would like to 
thank Arjun Dey for his help with the ``nod-and-shuffle'' sky subtraction
which proved critical in our analysis.

\end{document}